# Growth and organization of (3-Trimethoxysilylpropyl) diethylenetriamine within reactive amino-terminated self-assembled monolayer on silica

Yannick Dufil[a,b,*], Virginie Gadenne[a], Pascal Carrière[c], Jean-Michel Nunzi[a,b], Lionel Patrone[a,*]

[a] Aix Marseille Univ, Université de Toulon, CNRS, IM2NP, UMR 7334, Marseille, France, & Yncréa Méditerranée, ISEN Toulon, Maison du Numérique et de l'Innovation, Place G. Pompidou, 83000 Toulon, France

[b] Dpt. of Chemistry, Queen's University, 99 University Avenue, Kingston, ON K7L 3N6, Canada

[c] MAtériaux, Polymères, Interfaces et Environnement Marin - MAPIEM Laboratory, Université de Toulon, CS 60584, 83041 Toulon cedex 9, France

ABSTRACT

Alkane chains are the most commonly used molecules for monolayer fabrication. Long chains are used for their strong van der Waals interactions inducing good layer organization. Amine function-terminated alkyl chains are of great interest and are widely used for further surface functionalization. Since it is mandatory that such layers be organized to provide amine moieties at the surface, the present study deals with exploring amine-terminated SAM formation as an alternative to the usual aminopropylalkylsilane SAM. Additionally, using a long $NH_2$ terminated alkyl chain allows the formation of hydrogen bonding thanks to the two NH moieties born along the chain. Furthermore, such hydrogen bonding makes possible to shorten the molecule length while preserving a well-organized monolayer. For this purpose we performed a complete study of the grafting of (3-Trimethoxysilylpropyl) diethylenetriamine (DETAS) on native silicon oxide using various solvents, relative humidity and temperature values. Grafting kinetics was monitored by ellipsometry and goniometry, and SAM structure and organization using AFM and ATR-FTIR spectroscopy. Hydrogen bonding was evidenced within the SAM growth process and in the final complete SAM. We believe such study enables a better control of good quality DETAS SAM in order to improve their efficiency in further surface functionalization applications.

## 1. Introduction

Monolayer self-assembling is a convenient, simple and versatile technic to build ordered bi-dimensional nanoscale structures. It has been widely used in order to tailor interfacial properties, such as wettability or hydrophobicity, as well as the work-function of most materials. Self-assembling molecules possess three parts which can be modified to meet the required criteria: the head, the body, and the tail. By proper modification of the body of the molecule one can act on the ability of the molecule to form ordered monolayers. Particularly, organosilane-based self-assembled monolayers (SAMs) have attracted much attention for several years [1–18]. Those SAMs are known as being very stable and robust due to chemical bonding with the surface and siloxane cross-linking. Since they self-assemble on silicon dioxide they offer the possibility to build molecular devices compatible with silicon-based technology. Moreover, lots of commercial organosilane molecules with different terminal moieties make them very attractive to functionalize oxide surfaces. Among these molecules, 3aminopropyltrimethoxysilane (APTMS) is widely used to immobilize various entities via chemical bonding, including proteins [3], DNA [4], metal nanoparticles [5,6] or other specific molecules [7–9] such as fullerenes [9]. Most research groups prepared APTMS SAMs from a liquid phase using various solvents (water, ethanol, toluene, and so on) [3,5,9,12], and various grafting times [12–16]. However, APTMS very often leads to disordered multilayers [17]. Therefore, in this work we have been interested in an alternative amino-organosilane that could enable preparation of more ordered amine-terminated SAMs on a silicon surface. For this purpose, we have chosen to study the SAM formation and organization of (3-Trimethoxysilylpropyl) diethylene-triamine organosilane (DETAS). SAMs from such molecule have been studied for different applications such as immobilization of other organic species on top [19], controlling the diffusion of evaporated metals on the SAM [20], coating of silica nanoparticle for water treatment without organic solvent [21], patterning by using DETAS self-assembled monolayer to promote cell growth at specific sites of the surface [22], or specific robust adsorption of PEDOT nanofilms [23]. For such possible applications, it is important that DETAS SAMs should be well organized. Regarding SAM formation, such organosilane is bearing a longer alkyl chain that enables a higher flexibility to promote molecular interaction, and including also two NH moieties capable of introducing structuring hydrogen bonding between neighboring molecules in the SAM.

## 2. Experimental details

### 2.1. Chemicals

DETAS [(3-Trimethoxysilylpropyl) diethylene-triamine] ([MeO]$_3$-Si-C$_3$[NH-C$_2$]$_2$-NH$_2$) (Fig. 1), acetone, methanol, ethanol, iso-propanol, butanol, toluene, dichloromethane, trichloromethane, hydrogen peroxide 30% (H$_2$O$_2$), and sulfuric acid (H$_2$SO$_4$) are of synthesis grade and absolute ethanol of analytical grade, and all products were used with no further purifications. The water used in all experiments was deionized and purified by a Millipore system and its resistance was ~18 MΩ.

### 2.2. Sample cleaning and SAM preparation

Si substrates of ~1 cm by 1.5 cm were degreased in three consecutive ultrasonic baths of 10 min each subsequently with acetone, dichloromethane, and iso-propanol. The samples were then transferred into a mixture of 3 volumes of H$_2$O$_2$ and 7 volumes of H$_2$SO$_4$ heated at 110 °C for 30 min to remove the last traces of organic pollutant on the surface. After that, the samples were thoroughly rinsed with deionized water and stored into a water beaker upon usage for no more than few minutes. The samples were then dried under a nitrogen flow and quickly plunged into a DETAS solution. The temperature was controlled with an oil bath on a thermostated plate by a Ministat from Huber, inside a glove box (818-GB from Plas-Lab) filled with grade 4.5 nitrogen. Atmospheric relative humidity was monitored (using a TFA Dostmann 30.5013). Relative humidity was controlled with desiccants and a heating system. After SAM preparation, samples were rinsed with clean solvent used for grafting, and treated with ultrasonic bath of the same solvent for 5 min. After that, samples were dried under nitrogen flow and characterized right away.

### 2.3. SAM characterization

Ellipsometric measurements were carried out on a Sentech SE400 at 70° angle with a 632.8 nm laser. Parameters used were 3.8750 and 0.0018 for the real and imaginary indices of the silicon substrate, 1.46 and 0 for the native oxide layer, and 1.45 and 0 for the indices of DETAS with an expected thickness of 14 Å [24]. Thickness of the native silicon oxide layer was recorded on control samples that underwent the same cleaning process as the grafted samples. At least five different points of the sample surface were measured, possibly on different samples, in order to give a mean value with error bars amplitude between the minimum and the maximum values measured. Goniometric measurements were conducted with a Kruss drop shape DSA 10 MK2 analyzer, with deionized water as liquid. The sample was cleaned under nitrogen flow after measuring. AFM measurements were performed in Tapping mode with 40 N/m force constant silicon tips, using a Bruker Multimode 8 equipped with a Nanoscope V electronics. Fourier transformed infrared spectra were recorded on a Thermofisher IS50 spectrometer in single reflection ATR mode at an incidence angle of 45° on diamond. 96 scans were accumulated in a spectral range within 4000–400 cm$^{-1}$ with a resolution of 4 cm$^{-1}$. Background spectrum of the bare silicon substrate was recorded in the same conditions and subtracted.

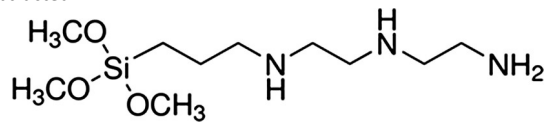

**Fig. 1.** Schematic representation of (3-Trimethoxysilylpropyl) diethylene-triamine (named "DETAS"), approximately 14 Å in length.

## 3. Results and discussion

### 3.1. Growth kinetics of SAMs of DETAS

In order to identify the time-scale necessary to reach a complete single monolayer coverage of DETAS on silica, the SAM growth kinetics was studied with a concentration of 5.10$^{-2}$ mol·l$^{-1}$ in different solvents under 33 ± 5% relative humidity. High temperature (31.5 ± 1.5 °C) grafting was conducted in alcohol solvents and lower temperature (11.0 ± 0.5 °C) in toluene. Grafting durations chosen for kinetics study were 1 s, 5 s, 10 s, 30 s, 1 min, 5 min, 10 min, 1 h, 3 h and 24 h. Time evolution of the thickness is plotted in Fig. 2 for the different growth conditions.

All experimental data could be well fitted using a simple Langmuir type growth equation:

$$Th = A \exp(-t/\tau) + Th_{max} \tag{1}$$

Where Th stands for the thickness, A coefficient arises from a delay in the grafting process, $\tau$ is the time constant and Th$_{max}$ is the thickness reached at the end of the grafting. Values of the fitting parameters are displayed in Table 1. As can be seen, the thickness jumps from 0 to about 8 Å at t = 0 s, upon dipping in the solution. This maybe an indication that DETAS molecules form a film at the surface of the solution that is deposited on the surface while dipping, thus forming a pseudo Langmuir-Blodgett film on the sample surface.

As can be seen in Table 1, the nature of the solvent and the grafting temperature seem to play very little role in the growth kinetics since the growth time constants and delay coefficients are comparable within confidence limits.

Growth time constant values are in the range of 7000–10000 s (~1.9–2.8 h) and the final SAM thickness is obtained after ~30000 s, i.e., ~8 h. The growth kinetics study of DETAS SAM was performed by Demirel et al. [24] at a concentration of ~4.10$^{-3}$ M in ethanol.[1] Authors found that DETAS SAM reached a maximum thickness of ~15 Å after ~12 h grafting. Given that in our experiments the concentration is about ten times higher, we should expect a growth process ten times quicker. But as shown in a previous study on alkylsilane growth [25], the temperature and the humidity both play a crucial role in the SAM growth kinetics. No information is given in the article by Demirel et al. [24] on the relative humidity conditions. However, as they worked at ambient temperature, our temperature around 31.5 °C was higher, which was shown to slower the SAM growth with favoring a more disordered phase that hinders molecule diffusion.

The first steps of the growth are visible in the semi-logarithm time scale presented in the insets of Fig. 2. As can be seen, the thickness remains at about 8 Å during ~2000 s in the same manner as in the growth analysis by Demirel et al. [24]. It may indicate a first step of growth were the molecules are lying bent on the surface before a possible interplay of both van der Waals bonds between methylene groups and hydrogen bonds between the amino NH and NH$_2$ moieties. This is further supported by the quicker increase that we observe for the water contact angle than for the thickness, as in Ref. [24] (Fig. 3).

---

[1] The concentration used in reference 24 is 1% in volume ratio. With a density of 1.03 for DETAS it corresponds to a mass concentration of ~10$^{-2}$ g/ml which gives a molar concentration of ~4.10$^{-3}$ M with a molar weight of 265 g.

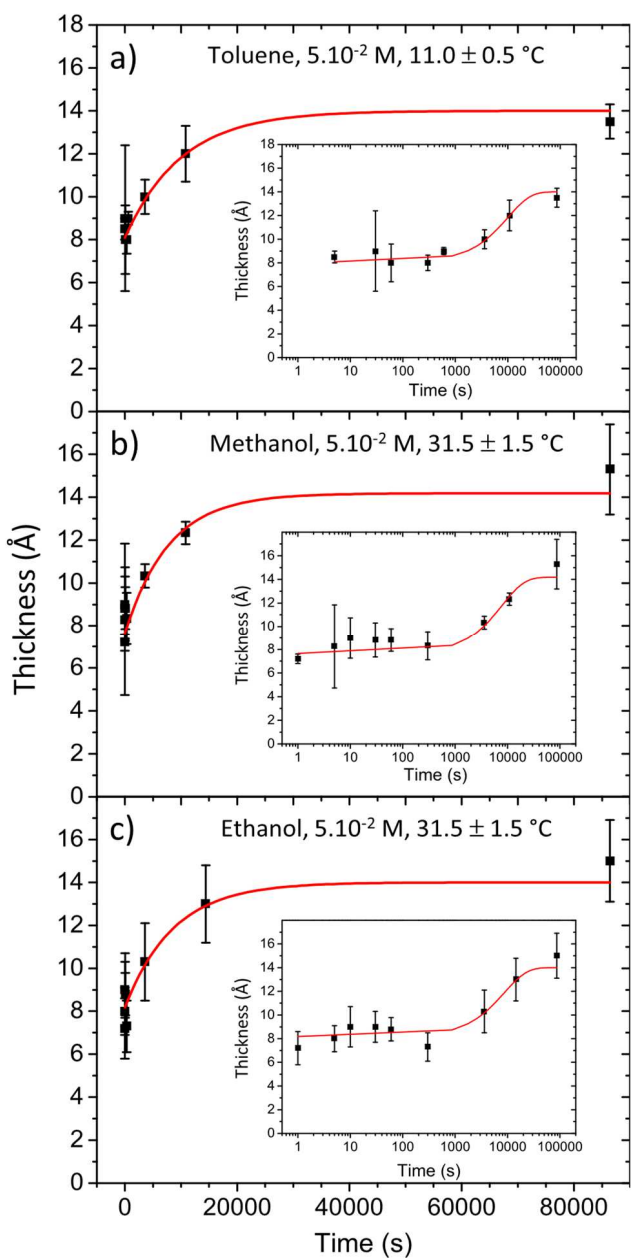

**Fig. 2.** Grafting kinetics curves of DETAS self-assembled monolayer at $5.10^{-2}$ M and under $33 \pm 5\%$ RH in: (a) toluene at $11.0 \pm 0.5$ °C, (b) methanol at $31.5 \pm 1.5$ °C, (c) ethanol at $31.5 \pm 1.5$ °C. Inset shows the curve with logarithm time scale to highlight the early stage of the growth.

**Table 1**
Fitting values of the growth kinetics of SAM of DETAS in different solvents and temperature.

| Solvent | Temperature (°C) | A | $\tau$ ($10^3$ s) | $Th_{max}$ (Å) |
|---|---|---|---|---|
| Toluene | $11.0 \pm 1.0$ | $-5.9 \pm 0.3$ | $10.6 \pm 3.6$ | 14 |
| MeOH | $31.5 \pm 1.0$ | $-5.8 \pm 0.3$ | $8.6 \pm 3.9$ | 14 |
| EtOH | $31.5 \pm 1.0$ | $-6.4 \pm 0.3$ | $7.4 \pm 1.2$ | 14 |

Indeed, as can been seen, compared to the first order Langmuir type fitting function of the thickness time evolution, water contact angle values are positioned at lower time within the first 200 min of SAM growth. This can be explained by molecules lying flatter on the surface at the early stage of growth (t < 200 min), which means the water droplet probes a mixture made of methylene $CH_2$ groups which contribute to increase the water contact angle, with both more hydrophilic terminal amino $NH_2$ moieties and uncovered oxide surface.

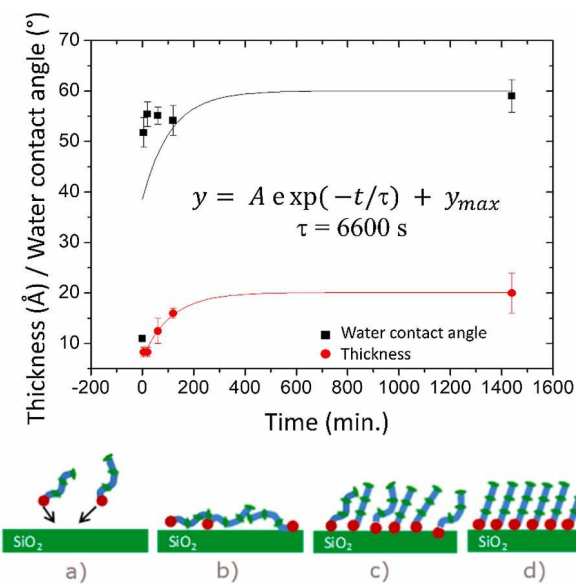

**Fig. 3.** Top: Grafting kinetics of DETAS self-assembled monolayer at $10^{-2}$ M, 30 °C, under 80% RH in toluene for the thickness (filled black squares) and water contact angle (filled red circles). Fitting curves using a first order Langmuir type law was obtained from thickness data. Bottom: scheme of the proposed grafting process with: (a) molecules are adsorbed towards the surface, (b) molecules are lying flat on the surface, (c) molecules are beginning to stand up at the surface, (d) final organized compact SAM. (For interpretation of the references to colour in this figure legend, the reader is referred to the web version of this article.)

On the contrary, at the end of the growth, water contact angles are well measured on $NH_2$ groups, with an end value of ~60° compatible with the values found in the literature for $NH_2$ terminated SAMs [26] and particularly with the value of 65° reported for the final DETAS SAM [24]. Fig. 2 shows complete SAMs were obtained after at least 8 h whatever the solvent, temperature and humidity. Therefore, in the following, to warrant a complete SAM coverage we fixed the grafting duration to 24 h.

### 3.2. Effect of the relative humidity

As the importance of dipping time and concentration is quite wellknown, we first investigated the effect of relative humidity (RH) on the formation of self-assembled monolayers of DETAS on native silicon oxide. For this purpose, among possible solvents for DETAS we chose to use toluene rather than alcohols because its non-polar nature makes it more sensitive to the water content in solution. The effect of relative humidity on SAM formation was tested from 25% RH up to 80% RH with a DETAS solution in toluene at $1.10^{-2}$ mol·$l^{-1}$ at room temperature ($22 \pm 2$ °C) and the dipping time was set to 24 h. The SAM thickness, measured by ellipsometry, was plotted versus relative humidity in Fig. 4.

Contact angle measurements were also recorded but are irrelevant with using toluene as a solvent as we have noticed that dipping a substrate into toluene is enough to considerably modify its contact angle with water (from 10° to 30° depending of the surface), unless it is cleaned in ultrasonic bath followed by hot acid bath. As can be seen in the graph, a SAM with about the desired thickness of 14 Å [24] is achieved provided that relative humidity is kept under ~40%. When relative humidity gets higher, the thickness increases by around two orders of magnitude which can be explained by multilayer formation. The sudden thickness increase above 40% RH indicates a change in the behavior of DETAS molecule in solution at this threshold value. Our hypothesis is that DETAS behavior in non-polar solvent is analogous to the behavior of surfactants.

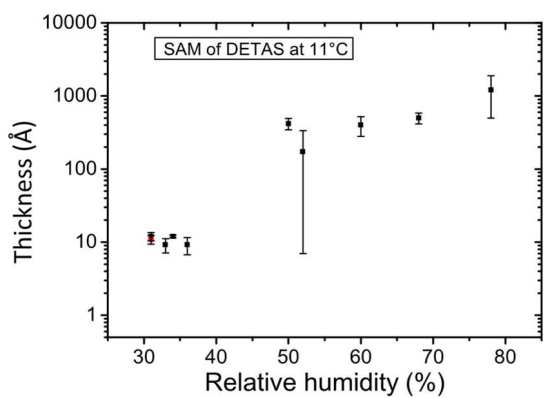

**Fig. 4.** Effect of relative humidity on the thickness of self-assembled monolayers of DETAS prepared on silica from toluene solution (1.10$^{-2}$ M, 22 ± 2 °C).

As relative humidity increases, water dissolves in toluene until it reaches equilibrium allowing water to regroup around DETAS molecules forming micelle-like structures. Then these reverse micelles can react to form oligomers that are transferred to the surface before chemisorption. A visible layer was observable on each interface with the solution (solution-air, solution-substrate, and solution-beaker) which further confirms the surfactant-like behaviour of these structures. A foam was also observable in the solution for higher concentration of DETAS which is consistent with an oligomerized micelle-like behavior. AFM analysis was conducted on samples at high relative humidity at different grafting times, revealing cauliflower-like structures growing on the substrate at the beginning of the grafting and a superposition of many layers of oligomerized reverse micelles at the end of the grafting (Fig. 5). Therefore, in the following of the study we worked at low humidity to avoid such reverse micelle superimposition in the final SAM.

*3.3. Low humidity Self-assembled monolayers*

Regarding low relative humidity (< 40%) grafting, smooth (0.28 nm average roughness) monolayers about 14 Å thick were obtained as confirmed by ellipsometry together with AFM profiling by denting the self-assembled monolayer with a soft fabric. AFM image of DETAS self-assembled monolayer grafted with low relative humidity in toluene and ethanol (Fig. 6) exhibits similar smooth layers with very few defects. Therefore, using alcohol as solvent will be preferred in the following since such solvent in which DETAS is better solubilized are less sensitive to the relative humidity than toluene. Some white spots of about 1–30 nm remain on the layers, which can be attributed to some oligomers that have formed either on the surface or in solution. Lowering further the relative humidity may improve the monolayer quality.

Siloxane monolayer organization is usually driven by van der Waals interaction between long alkyl chains. This process is well known and leads to remarkable self-assembled monolayers provided the alkyl chain is long enough [2,25]. DETAS however does possess two secondary amine groups inside the alkyl chain alongside with a primary amine group at the head of the molecule. Those amine groups are able to form hydrogen bonds inside the layer to stabilize more efficiently the monolayer. Hydrogen bonds are stronger than van der Waals interaction by several orders of magnitude, allowing better organization of the monolayer than pure alkane chains with comparable short length. The ATR-FTIR spectra of a self-assembled monolayer of DETAS are shown in Fig. 7 in different spectral regions.

In Fig. 7a, one can see the two doublet peaks positioned at mean values of ~3712 and ~3612 cm$^{-1}$ that are assigned to hydrolyzed methoxysilane groups that have not crosslinked in the usual condensation process [27]. We can note that at the base of these two doublet peaks, there is a broad band which seems to indicate that some of those silanol groups are hydrogen bonded. The band at 3406 cm$^{-1}$ and the large one within 3300–2800 cm$^{-1}$ that could be decomposed into three Lorentzian bands at 3264, 3142 and 3017 cm$^{-1}$ are assigned to the N-H stretch of primary and secondary amines [28–30]. The band at 3264 cm$^{-1}$ can be a consequence of the shift to lower frequency of the 3406 cm$^{-1}$ band for the secondary amines that experience hydrogen bonding in the SAM. The other two bands of the primary amine stretch at 3142 and 3017 cm$^{-1}$ are forming a broad band and indicate that there is hydrogen bonding within the self-assembled monolayer. Fig. 7b exhibits two bands at 1568 and 1494 cm$^{-1}$ that can be attributed to amine group deformation. Usually such modes appear at higher frequencies but again such a shift towards lower frequencies may be due to hydrogen bonding [31]. The extent of such shift depends on the intensity of the hydrogen bonding. For example, Chiang *et al.* [32] assigned the band at 1561 cm$^{-1}$ for aminopropyltrimethoxisilane SAM deposited on silica gel to NH$_2$ deformation modes with strong hydrogen bonding.

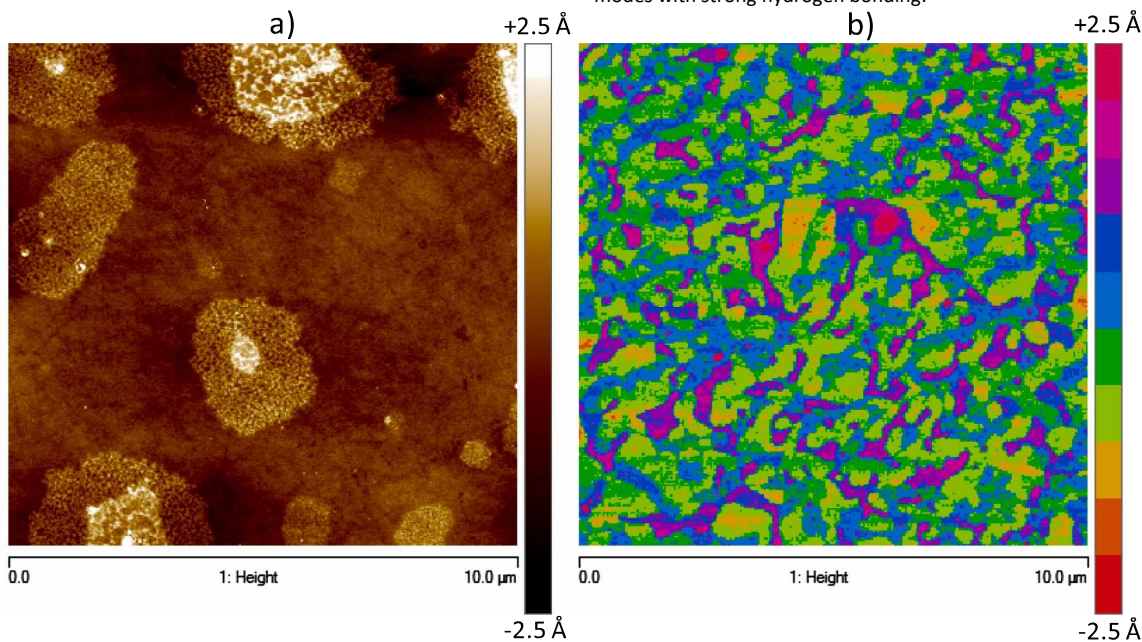

**Fig. 5.** AFM images (10 × 10 μm$^2$) showing: (a) cauliflower structures growing in the first stage of grafting, and (b) reverse micelle deposition in the later stage of grafting in toluene. Grafting conditions were: 10$^{-2}$ M of DETAS in toluene, ambient temperature, high humidity > 40%.

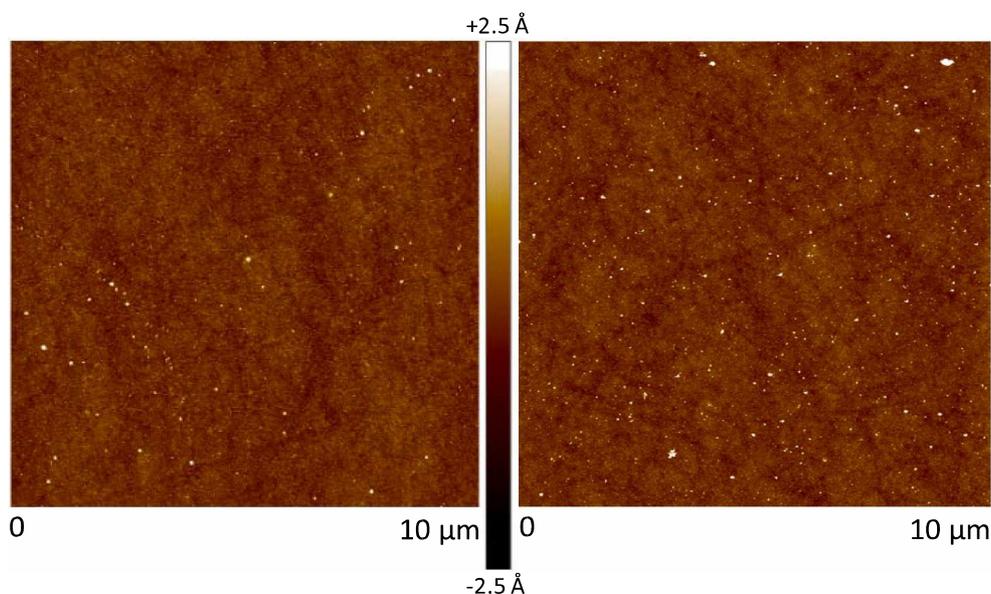

Fig. 6. AFM image of DETAS self-assembled monolayer: (a) $5.10^{-2}$ M in ethanol, 26 °C and 20% RH; (b) $10^{-2}$ M in toluene, 24 °C and 23% RH.

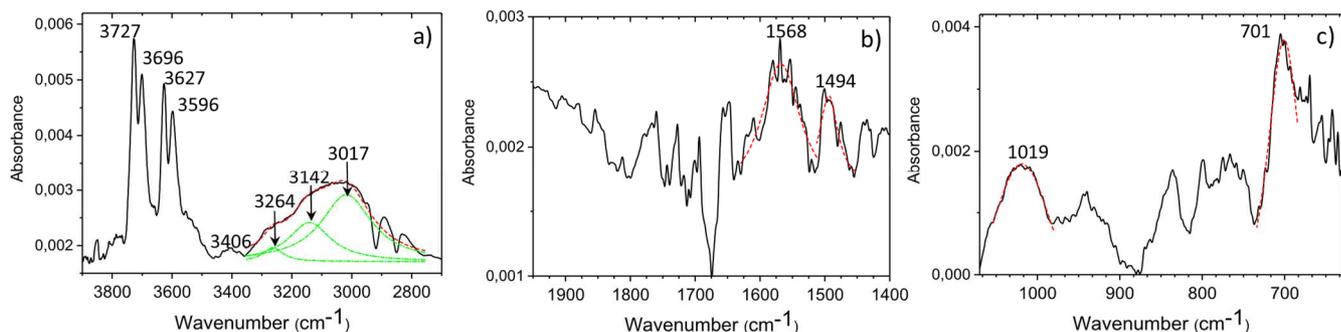

Fig. 7. ATR-FTIR spectra of DETAS self-assembled monolayer on silicon substrate (prepared from a solution $5.10^{-2}$ M in EtOH, 26 °C and 20% RH). Lorentzian fits have been performed (dashed green and red curves).

### 3.4. Carbon chain length and concentration of alcohols on grafting

Effects of concentration on the self-assembled monolayer was also tested for DETAS concentrations from $10^{-3}$ to $5.10^{-1}$ mol·l$^{-1}$ in three alcohol solvents that were shown to be less sensitive to relative humidity than toluene, certainly because of their high polarity. No micellar behavior should appear due to the high solubility of water and DETAS molecules in alcohols. Relative humidity was around 41%, temperature 33 °C and grafting time was 24 h for all samples. The thickness, measured by ellipsometry, plotted versus DETAS concentration is given in Fig. 8 for each of the three solvents used.

As can be seen in the graph, low concentration below $10^{-2}$ mol·l$^{-1}$ leads to incomplete self-assembled monolayers whereas higher concentration around $5.10^{-1}$ mol·l$^{-1}$ leads to SAMs with the expected thickness, but presenting high deviation in the thickness measurement, which indicates that a lot of defects or adsorbates are present at the SAM surface. A minimum concentration of $2.5.10^{-2}$ mol·l$^{-1}$ seems to be needed to achieve a good quality compact layer. Concentration can be further increased up to $5.10^{-2}$ mol·l$^{-1}$ without damaging the monolayer quality. It is also noticeable that the longer the alcohol chain, the better the quality of the monolayer, suggesting that the alkane chain is playing a role in the organization of the self-assembled monolayer. Similarly, the addition of long alkane chain solvent like hexadecane can improve organization and quality of octadecyltrichlorosilane SAMs [25].

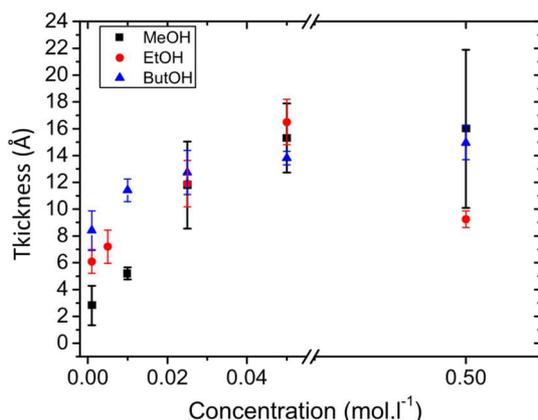

Fig. 8. Influence of both concentration and alcohol carbon chain length on the thickness of DETAS SAM prepared at 33 °C and 41% RH. Grafting duration is 24 h.

In Fig. 7c, the peak at 701 cm$^{-1}$ is assigned to N-H swaging on both primary and secondary amines [33]. Furthermore, formation of siloxane bonds can be inferred from the band located at 1019 cm$^{-1}$ [34,35].

## 4. Conclusion

Amino-terminated self-assembled monolayers are currently widely used in both industrial and research-oriented applications. We demonstrated that relative humidity plays a key role in the formation of DETAS SAMs with non-polar solvent like toluene and leads to multilayers through micellar deposition when it reaches 40% and above. Observation suggests that for obtaining a good DETAS SAM quality, a minimal DETAS concentration of $2.5 \times 10^{-2}$ M together with a long chain of alcohol solvent are needed. Kinetics show an immediate molecule adsorption within the very first moment of the SAM formation process which is uncommon for organosilane grafting. Thickness remains almost constant for more than 30 min before increasing with the molecule organization through hydrogen bonding between primary ($NH_2$) and secondary (NH) amine moieties, which was evidenced by infrared spectroscopy. Further steps are in progress concerning the functionalization to immobilize species on top of DETAS SAMS.

## Authorship contribution statement

**Yannick Dufil:** Methodology, Investigation, Validation, Writing original draft. **Virginie Gadenne:** Investigation, Supervision, Writing review & editing. **Pascal Carrière:** Investigation, Data curation. **JeanMichel Nunzi:** Supervision, Project administration, Funding acquisition. **Lionel Patrone:** Supervision, Conceptualization, Formal analysis, Project administration, Funding acquisition, Writing - original draft.

## Declaration of Competing Interest

The authors declare that they have no known competing financial interests or personal relationships that could have appeared to influence the work reported in this paper.

## Acknowledgements

FFCR-FCRF "New collaboration programme" between France and Canada is acknowledged for financial support through the "SAMOS" project, and particularly for funding the PhD of Y. Dufil. Funding from PHC Barrande 2018 (40672RL), Campus France, Ministères de l'Europe et des Affaires étrangères (MEAE) et de l'Enseignement supérieur, de la Recherche et de l'Innovation (MESRI), as well as financial support from Ministry of Education, Youth and Sports, within the projects LTC17058 and 8J18FR011, and by COST Action CA15107 MultiComp are also acknowledged. Y.D., V.G., and L.P. thank P. Fitl, J. Lančok, D. Tomeček and J. Vlček. J.-M.N. thanks CNRS and Aix-Marseille Université for funding his invitation at IM2NP. Equipment was mainly funded by the "Objectif 2" EEC program (FEDER), the "Conseil Général du Var" Council, the PACA Regional Council, Toulon Provence Méditerranée and ISEN-Toulon which are acknowledged.